\documentclass[letterpaper,10 pt,journal,twoside]{IEEEtran}

\usepackage{cite}
\usepackage{amsmath,amssymb,amsfonts}
\usepackage{algorithmic}
\usepackage{graphicx}
\usepackage{textcomp}
\usepackage{xcolor}
\usepackage{multirow}

\usepackage{tikz}
\usepackage{textcomp}
\usepackage{hyperref}

\newcommand\copyrighttext{
  \small \textcopyright 2020 IEEE. Personal use of this material is permitted. Permission from IEEE must be
  obtained for all other uses, in any current or future media, including reprinting/republishing this material for advertising or promotional purposes, creating new collective works, for resale or redistribution to servers or lists, or reuse of any copyrighted component of this work in other works. \newline This article has been accepted for publication in  IEEE Control Systems Letters. Citation information: DOI \href{https://doi.org/10.1109/LCSYS.2020.3001490}{10.1109/LCSYS.2020.3001490}}
\newcommand\copyrightnotice{%
\begin{tikzpicture}[remember picture,overlay]
\node[anchor=north,yshift=0.01pt] at (current page.north) {\fbox{\parbox{\dimexpr\textwidth-\fboxsep-\fboxrule\relax}{\copyrighttext}}};
\end{tikzpicture}%
}

\def\BibTeX{{\rm B\kern-.05em{\sc i\kern-.025em b}\kern-.08em
    T\kern-.1667em\lower.7ex\hbox{E}\kern-.125emX}}

\newcommand{\real}{\mathcal{R}}
\newcommand{\vo}[1]{\boldsymbol{#1}}
\newcommand{\mo}[1]{\boldsymbol{#1}}
\newcommand{\A}{\vo{A}}
\newcommand{\B}{\vo{B}}
\newcommand{\x}{\vo{x}}
\newcommand{\y}{\vo{y}}
\newcommand{\w}{\vo{w}}
\newcommand{\n}{\vo{n}}

\newcommand{\mub}{\vo{\mu}}
\newcommand{\sigb}{\vo{\Sigma}}

\newcommand{\xdot}{\dot{\vo{x}}}

\newcommand{\param}{{\vo{\Delta}}}

\newcommand{\pdfp}{p(\param)}
\newcommand{\set}[1]{\mathcal{#1}}
\newcommand{\Exp}[1]{\mathbb{E}\left[#1\right]}
\newcommand{\E}[1]{\Exp{#1}}
\newcommand{\basis}[2]{\phi_{#1}\left(#2\right)}
\newcommand{\F}{\vo{F}}
\newcommand{\I}[1]{\vo{I}_{#1}}

\newcommand{\K}{\vo{K}}

\newcommand{\R}{\vo{R}}
\newcommand{\Q}{\vo{Q}}

\newcommand{\Phin}[1]{\vo{\Phi}_{#1}(\param)}
\newcommand{\Phint}[1]{\vo{\Phi}^T_{#1}(\param)}

\newcommand{\Thp}{\vo{\Theta}(\param)}
\newcommand{\Thpt}{\vo{\Theta}^T(\param)}
\newcommand{\Thnp}[1]{\vo{\Theta}_{#1}(\param)}
\newcommand{\Thnpt}[1]{\vo{\Theta}^T_{#1}(\param)}

\renewcommand{\vec}[1]{\textbf{vec}\left(#1\right)}

\newcommand{\inner}[1]{\left\langle \vo{e}\phi_i\right\rangle} 

\newcommand{\eqnlabel}[1]{\label{eqn:#1}}

\newcommand{\eqn}[1]{(\ref{eqn:#1})}

\DeclareMathAlphabet{\mathbfsf}{\encodingdefault}{\sfdefault}{bx}{n}

\newcommand{\C}{\vo{C}}

\newcommand{\Ap}{\A(\param)}
\newcommand{\Apt}{\A^T(\param)}
\newcommand{\Bp}{\B(\param)}

\newcommand{\Bpt}{\B^T(\param)}

\newcommand{\domain}[1]{\set{D}}

\newcommand{\elab}[1]{\label{eqn:#1}}

\newcommand{\etal}{\textit{et. al. \,}}

\begin{document}
\title{\LARGE \bf Robust Kalman Filtering with Probabilistic Uncertainty in System Parameters
\thanks{This work was supported by NSF Grant \#1762825.}}

\author{Sunsoo Kim$^{1,4}$, Vedang M. Deshpande$^{2,4}$, and Raktim Bhattacharya$^{3,4}$
\thanks{$^{1}$Sunsoo Kim is a Ph.D student in the Department of Electrical and Computer Engineering. Email: {\tt\small kimsunsoo@tamu.edu}}%
\thanks{$^{2}$Vedang M. Deshpande is a Ph.D student in the Department of Aerospace Engineering. Email: {\tt\small vedang.deshpande@tamu.edu}}
\thanks{$^{3}$Raktim Bhattacharya is with the Faculty of Aerospace Engineering. Email: {\tt\small raktim@tamu.edu}}%
\thanks{$^{4}$Texas A\&M University, College Station, TX 77843, USA.}}

\maketitle
\thispagestyle{empty} % Removes the page number in the first
\copyrightnotice

\begin{abstract}
In this paper, we propose a robust Kalman filtering framework for systems with probabilistic uncertainty in system parameters. We consider two cases, namely discrete time systems, and continuous time systems with discrete measurements. The uncertainty, characterized by mean and variance of the states, is propagated using conditional expectations and polynomial chaos expansion framework. The results obtained using the proposed filter are compared with existing robust filters in the literature.  The proposed filter demonstrates better performance in terms of estimation error and rate of convergence.
\end{abstract}


\section{Introduction}
\IEEEPARstart{R}{obust}
 filtering algorithms such as $\mathcal{H}_{2}/\mathcal{H}_{\infty}$ filters and robust Kalman filters, have been developed to address uncertainty in system models. In the $\mathcal{H}_{2}$/$\mathcal{H}_{\infty}$ framework, filters are designed to minimize the impact of exogenous signals, i.e. process and sensor noise, on the estimation error \cite{geromel2002robust, lacerda2011robust, duan2013lmis, green2012linear, lewis2017optimal}. A robust Kalman filter is an extension of the well known Kalman filter, which can handle uncertainties in the system \cite{xie1994robustDT}. In this framework, the filter is designed to minimize an upper bound on the estimation error variance \cite{zhu2002design,yang2002robust,liu2018robust,abolhasani2018robust,xie1994robustCT,shi1999robust}, or the worst-case error variance \cite{sayed2001framework,zorzi2016robust,zorzi2018robust}. Our work falls in the category of robust Kalman framework.

Existing robust Kalman filter algorithms can be categorized based on how system uncertainty is represented, which is  assumed to be parametric. The uncertainty is either represented as norm bounded parameter uncertainty \cite{xie1994robustDT, zhu2002design, yang2002robust, liu2018robust, abolhasani2018robust, xie1994robustCT, shi1999robust}, or polytopic parametric uncertainty \cite{shaked2001new, lewis2017optimal}. In this work, we model parametric uncertainty as random variables with known probability density function (PDF). To the best of our knowledge, this is the first work on robust Kalman filtering with probabilistic system uncertainty.

We present two robust Kalman filtering algorithms with probabilistic uncertainty in system parameters.
The first algorithm is for discrete-time (DT) system where the dynamics and measurements are both in discrete time. The second algorithm is for continuous-time (CT) dynamical systems with discrete-time measurements. In both these cases, mean and variance of uncertain states are calculated using a formulation based on conditional expectation. For the CT system, we apply polynomial chaos (PC) framework  {which provides a deterministic and computationally tractable approach to propagate the uncertainty.}

The rest of the paper is organized as follows. We first present the problem formulation with uncertainty in CT and DT domain in \S \ref{sec:Problem Formulation} followed by a discussion on polynomial chaos framework in \S \ref{sec:Polynomial Chaos Theory}. \S \ref{sec:Robust Kalman filter} presents the proposed robust filter. Simulation results are presented in \S \ref{sec:result} followed by concluding remarks in \S \ref{sec:Conclusion}.

\section{Problem Formulation} \label{sec:Problem Formulation}
The objective of filtering is to estimate the state-trajectory $\x(t)$ or $\x_k$ of a physical process in CT or DT, given noisy measurements. The uncertainty in the system parameters, in the external excitation (process noise), and in the measurement errors  (sensor noise), are all treated as probabilistic.  The model for the evolution of the state is assumed to be the following linear-time-varying \textit{stochastic} system,
\begin{subequations}
\begin{align}
\text{CT:} \  & {\xdot(t) = \A(\param_{k-1})\x(t) + \B(\param_{k-1})\w(t)}, \elab{sys_ct0} \\
% \nonumber \\ & \quad \quad \quad {t_{k} \leq t < t_{k+1}, \elab{sys_ct0}} \\
\text{DT:} \ & {\x_{k} = \A(\param_{k-1}) \x_{k-1} + \B(\param_{k-1}) \w_{k-1}}, \elab{sys_dt0}
\end{align}
\elab{sys0}
\end{subequations}
{where $t_{k-1} \leq t < t_{k}$ for CT model \eqn{sys_ct0}.  $\x,\x_k\in\mathcal{R}^n$ represent the state vector, $\w,\w_k\in\real^m$ are  zero mean Gaussian noise processes with covariance $\Exp{\w(t)\w^T(\tau)} = \Q\delta(t-\tau)$ and $\Exp{\w_i\w^T_j} = \Q\delta_{ij}$ respectively, where $\delta(\cdot)$ and $\delta_{ij}$ are delta function and  Kronecker delta respectively.}

$\A(\cdot): \real^{d}\mapsto\real^{n\times n}$ and $\B(\cdot): \real^{d}\mapsto\real^{n\times m}$ are system matrices {with given functional dependence on $\param_{k}$}. The random vector $\param_{k}\in \real^{d}$ represents the uncertain parameters in the system matrix.
{In DT model \eqn{sys_dt0}, the parameter vector $\param_{k}$ is sampled at every time step. And in CT model \eqn{sys_ct0}, $\param_k$ is sampled at discrete time instants $t_k$, and its realization does not change within the time span $[t_{k},t_{k+1})$. In both cases, the sequence $\param_0, \param_1, \param_2, \cdots$ is assumed to be an independent and identically distributed random process with a given PDF.}

We also assume, the initial state for \eqn{sys0} is a random variable with a given PDF that is independent of the process noise $\w(t)$ or $\w_k$, and the system parameters $\param_{k}$.
%If $\x(t,\param,\w)$ or $\x_k(\param,\w)$ is the solution of \eqn{sys0}, we assume the model to be accurate enough such that the true states are sample paths of $\x(t,\param,\w)$ and $\x(k,\param,\w)$, respectively.

Measurement from sensors is modeled as
\begin{align}
 \y_k = \C \x_k + \n_k,\elab{y0}
\end{align}
which maps the state $\x_k$ to the output space $\y_k$ and is corrupted by sensor noise $\n_k$. In the output model, $\C\in\real^{m\times n}$ is deterministic and {$\n_k$ is zero mean Gaussian white noise with $\Exp{\n_i\n^T_j} = \R\delta_{ij}$.}
The process and sensor noise are assumed to be uncorrelated with {known $\vo{Q}$ and $\vo{R}$.}
%The variances of $\w$ and $\n$ are assumed to be {known and denoted by $\vo{Q}$ and $\vo{R}$ respectively.}

The objective here is to determine the \textit{unbiased} estimate of $\x(t)$ or $\x_k$ with \textit{minimum error-variance}, using the model defined by \eqn{sys0} and \eqn{y0}. This is achieved by extending the formulation for standard Kalman filtering, to systems with probabilistic uncertainty in system parameters, which is discussed in \S \ref{sec:Robust Kalman filter}. However before that, we briefly discuss the polynomial chaos framework that is used for propagation of uncertainty in CT systems.

\section{Polynomial Chaos Theory}\label{sec:Polynomial Chaos Theory}
Polynomial chaos is a deterministic framework to determine the evolution of
a stochastic process $\vo{\xi}(t,\param)$, where $\param\in\mathcal{D}_{\param} \subset\real^d$ represents the parameter space with known PDF $\pdfp$. Differential equations with probabilistic parameters e.g.
\begin{equation}
\vo{\dot{\xi}}(t,\param) = \vo{F}(t,\vo{\xi}(t),\param), \eqnlabel{autoDyn}
\end{equation}
are examples of such stochastic processes that are amenable for analysis using polynomial chaos theory.
Assuming $\vo{\xi}(t,\param)$ to be a second-order process, it can be expanded, with $\set{L}_2$ convergence \cite{CameronMartin,Xiu}, as
$$
\vo{\xi}(t,\param) = \sum_{i=0}^\infty \vo{\xi}_i(t)\phi_i(\param),
$$
where $\vo{\xi}_i(t)$ are time varying coefficients, and $\phi_i(\param)$ are known basis polynomials. For exponential convergence, $\phi_i(\param)$ are chosen to be orthogonal with respect to the PDF $p(\param)$, i.e.
\begin{equation*}
\Exp{\basis{i}{\param}\basis{j}{\param}}:= \int_{\mathcal{D}_{\param}}{\basis{i}{\param}\basis{j}{\param} \pdfp
\,d\param}  = h_i\delta_{ij}, % \eqnlabel{basisFcn}
\end{equation*}
where $h_i:=\int_{\mathcal{D}_{\param}}{\phi_i^2\pdfp\,d\param}$.
For computational purposes, we truncate the expansion to a finite number of terms, i.e. the solution of \eqn{autoDyn} is approximated by the polynomial chaos expansion as
\begin{align}
	\vo{\xi}(t,\param) \approx \vo{\hat{\xi}}(t,\param) =  \sum_{i=0}^N \vo{\xi}_i(t)\basis{i}{\param}. \elab{approxSoln}
\end{align}
For a more compact representation of the ensuing expressions, we define $\mo{\Phi}(\param)$ to be
\begin{align}
\mo{\Phi}(\param) &:= \begin{bmatrix}\basis{0}{\param}, & \cdots, & \basis{N}{\param}\end{bmatrix}^T, \text{ and } \\
\Phin{n} &:= \mo{\Phi}(\param)\otimes \I{n},
\end{align}
where $\I{n}\in\real^{n\times n}$ is identity matrix. We define matrix $\vo{\Xi}\in\real^{n\times(N+1)}$, with polynomial chaos  {coefficients $\xi_i$}, as $\vo{\Xi} := \begin{bmatrix} \vo{\xi}_0, & \cdots, & \vo{\xi}_N \end{bmatrix}.$
% \begin{align}
% \vo{\Xi} = \begin{bmatrix} \vo{\xi}_0, & \cdots, & \vo{\xi}_N \end{bmatrix}. \elab{defX}
% \end{align}
Therefore, $\vo{\hat{\xi}}(t,\param)$ can be written as
\begin{align}
\vo{\hat{\xi}}(t,\param) := \vo{\Xi}(t)\mo{\Phi}(\param) \eqnlabel{compactX}.
\end{align}
Noting that $\vo{\hat{\xi}} \equiv \vec{\vo{\hat{\xi}}}$, \eqn{compactX} becomes,
\begin{align}
\vo{\hat{\xi}} & \equiv  \vec{\vo{\hat{\xi}}} = \vec{\vo{{\Xi}}\mo{\Phi}(\param)} = \vec{\I{n}\vo{{\Xi}}\mo{\Phi}(\param)} \nonumber \\
 & = (\mo{\Phi}^T(\param)\otimes \I{n})\vec{\vo{{\xi}}}  =  \mo{\Phi}_n^T(\param)\vo{\xi}_\text{pc}, \eqnlabel{compactxpc}
\end{align}
where  $\vo{\xi}_\text{pc} := \vec{\vo{\Xi}}$, and $\vec{\cdot}$ is the vectorization operator.

The unknown coefficients $\vo{\xi}_\text{pc}$ are determined using one of many methods including Galerkin projection\cite{Ghanem1,ghanem:91}, stochastic collocation \cite{Marzouk2009a,NASA-CR-2003-212153}, and least-square minimization \cite{walters2003towards, hosder2007efficient}. In this work, we pursue the Galerkin projection approach to determine the coefficients $\vo{\xi}_\text{pc}(t)$ by first defining error $\vo{e}(t,\param):=\vo{\xi}(t,\param) - \mo{\Phi}_n^T(\param)\vo{\xi}_\text{pc}(t)$.
{The optimal coefficients $\vo{\xi}_\text{pc}(t)$  are then determined by setting projection of $\vo{e}(t,\Delta)$ against each basis to zero ensuring that the  error is orthogonal to the basis polynomials, i.e.}
\begin{align}
\int_\domain{\param} \vo{e}(t,\param) \phi_i(\param) p(\param) d\param = 0, \nonumber
\end{align}
for $i=0,\cdots,N$. {This results in a system of algebraic equations which can be solved for $\vo{\xi}_\text{pc}(t)$.}
If $\vo{\xi}(t,\param)$ is solution of a differential equation \eqn{autoDyn}, then the error is defined in terms of the equation error, as shown in \eqn{eqnerror}.

In general, polynomial chaos does not scale well with state-space and parameter dimension. The number of basis functions for a given order $r$ with $d$ independent random variables is $\frac{(d+r)!}{d!r!}$. With large number of parameters (increasing $d$), the number of basis functions, for a given order of approximation, will increase factorially and the computational cost will be prohibitive. This limits how large both $d$ and $r$ can be. Recent development in sparse polynomial chaos may scale better \cite{constantine2012sparse, conrad2013adaptive}. However, usually we can get quite good performance with low order approximations \cite{bhattacharyarobust, Fisher2009, bhattacharya2012linear, Dutta2010}. Unfortunately, the order of approximation, for which acceptable accuracy is achieved, has to be determined empirically.

For $d>1$, the polyvariate basis functions are determined from tensor-products of univariate polynomials, with limit on the total order of the product using Pascal's triangle, the univariate polynomials can be determined from different distributions. For a given distribution, using polynomials that are orthogonal with respect to the distribution, is usually chosen for exponential convergence \cite{Xiu}. Poor scalability of polynomial chaos is due to the tensor product of the basis functions. However, anisotropic tensor products \cite{babuska2004galerkin, babuvska2007stochastic} or anisotropic Smolyak cubature methods result in improved scaling \cite{nobile2008sparse}.

In this paper, we consider elements of $\param$ to be independent. However, in several applications this assumption may not valid. For such applications, suitable transformation such as Rosenblatt \cite{rosenblatt1952remarks}, Nataf \cite{der1986structural} and Box-Cox \cite{box1964analysis} transformation can be applied to arrive at a set of independent parameters. An overview of such techniques is described in the work by Elred \etal \cite{eldred2009comparison}.

\section{Robust Kalman filter}\label{sec:Robust Kalman filter}
In Kalman filtering, state estimation involves two steps: a) model-based uncertainty propagation to obtain the \textit{prior} state uncertainty, and b) incorporation of measurements to update the prior to \textit{posterior} state uncertainty by minimizing the error variance. With probabilistic uncertainty in the system parameters, along with process noise, the propagation step becomes complicated. In this paper, we solve this by computing the mean and variance of the propagated states using conditional expectations.

The new robust Kalman filtering algorithms, for uncertain DT and CT systems, are presented next.

\subsection{Discrete Robust Kalman Filter} \label{sec:DRF}
Let us consider the DT model given by \eqn{sys_dt0} and \eqn{y0} as
\begin{subequations}
\begin{align}
\x_{k} &= \A(\param_{k-1}) \x_{k-1} + \B(\param_{k-1}) \w_{k-1}, \elab{sys}  \\
\y_k &= \C\x_{k} + \n_k . \elab{y_dt}
\end{align}
\end{subequations}

\subsubsection{\textbf{Uncertainty propagation}} \label{sec:ddkf_prop}
The uncertainty in $\x_k(\param,\w)$, the solution of \eqn{sys}, is due to uncertainty in the initial condition $\x_0$, the uncertainty in the system parameters $\param_k$, and the process noise $\w_k$. % We will assume $\x_0$ and $\param$ are random variables with given probability density functions. For simplicity, we assume $\x_0$, $\param$, and $\w_k$ are statistically independent.
It is noteworthy, that due to the uncertainty in the system matrices, the PDF of state will not be Gaussian, even if $\x_0$ is Gaussian. However, we restrict ourselves to characterizing the first two moments of $\x_k(\param,\w)$ as defined below, since in this paper we are focusing on Kalman filtering. Let us define
\begin{subequations}
\begin{align}
\mub_k &:= \Exp{\x_k(\param,\w)}, \text{ and }\\
\sigb_k &:= \Exp{\left(\x_k(\param,\w)-\mub_k\right)\left(\x_k(\param,\w)-\mub_k\right)^T}.
\end{align}
\end{subequations}
Consequently, the propagation equation for $\mub_k$ is given by $$\mub_k = \E{\A(\param_{k-1})\x_{k-1}} + \E{\B(\param_{k-1})\w_{k-1}}.$$ We use the conditional expectation with respect to $\param_{k-1}$ to calculate the quantities in the previous equation. For a given $\param_{k-1}$, the propagation equations are similar to those in standard Kalman filter.
{Since, the distribution of $\x_0$ is given, and update step \eqn{disc_update1} has no uncertainty, it follows that the posteriors $\mub^{+}_{{k-1}}$ and $\sigb^{+}_{{k-1}}$ have no uncertainty, which is typical in robust filtering \cite{sayed2001framework,zorzi2016robust,zorzi2018robust}.} Therefore, we can write the propagation equation for conditional mean and variance as
% \begin{align}
% \mub^{-}_{k}(\param) &= \Ap\mub^{+}_{k}(\param), \elab{cond:mu}\\
% \sigb^{-}_{k}(\param) &= \Ap\sigb^{+}_{k}(\param)\Apt + \Q, \elab{cond:var}
% \end{align}
\begin{subequations}
\begin{align}
\mub^{-}_{k}(\param_{k-1}) &= \A(\param_{k-1})\mub^{+}_{{k-1}}, \elab{cond:mu}\\
\sigb^{-}_{k}(\param_{k-1}) &= \A(\param_{k-1})\sigb^{+}_{{k-1}}\A^T(\param_{k-1})  \nonumber \\ & \quad \quad + \B(\param_{k-1}) \Q \B^T(\param_{k-1}), \elab{cond:var}
\end{align}
\end{subequations}
{where $\mub^{-}_{k}(\param_{k-1})$ and $\sigb^{-}_{k}(\param_{k-1})$} are \textit{stochastic} since they depend on $\param_{k-1}$.
The \textit{total} mean and variance of {$\x_k(\param,\w)$} can be computed from the conditional mean and variance as
\begin{subequations}
\begin{align}
\mub^{-}_{k} &:= \E{\mub^{-}_{k}(\param_{k-1})}, \\
\sigb^{-}_{k}  &:= \E{\sigb^{-}_{k}(\param_{k-1})} + \textbf{Var}\left(\mub^{-}_{k}(\param_{k-1})\right).  \eqnlabel{totalStatVar}
%  \Exp{\left(\mub^{-}_{k}(\param)-\mub^{-}_{k}\right)\left(\mub^{-}_{k}(\param)-\mub^{-}_{k}\right)^T}.
\end{align}
\eqnlabel{totalStatDT}
\end{subequations}
With slight abuse of notation, we represent the conditional mean and variance as functions of $\param_{k-1}$, i.e. $\mub^{-}_{k}(\param_{k-1})$ and $\sigb^{-}_{k}(\param_{k-1})$. Whereas the total mean and variance are represented without the functional dependence, i.e. $\mub^{-}_{k}$ and $\sigb^{-}_{k}$.

Since the posterior $\mub^{+}_{{k-1}}$ is independent of $\param_{k-1}$, the total prior mean is calculated as
\begin{align}
\mub^{-}_{k} &:= \E{\mub^{-}_{k}(\param_{k-1})} = \E{\A(\param_{k-1})\mub^{+}_{k-1}} =\Bar{\A} \mub^{+}_{{k-1}}, \eqnlabel{mean_prior} %= \E{\Ap} \mub^{+}_{{k-1}} \nonumber\\ &=\Bar{\A} \mub^{+}_{{k-1}},
\end{align}
where $\Bar{\A}:=\E{\A(\param_{k-1})}$. The variance of conditional mean, $\textbf{Var}\left(\mub^{-}_{k}(\param_{k-1})\right)$, can be determined as
\begin{align}
&\textbf{Var}\left(\mub^{-}_{k}(\param_{k-1})\right) \eqnlabel{var_mu}  \\
& = \Exp{\left(\mub^{-}_{k}(\param_{k-1})-\mub^{-}_{k}\right)\left(\mub^{-}_{k}(\param_{k-1})-\mub^{-}_{k}\right)^T}\nonumber\\
&= \Exp{\left(\A(\param_{k-1})-\bar{\A}\right) \left(\mub^{+}_{k-1}  \mub^{+}_{k-1}{}^T \right) \left(\A(\param_{k-1})-\bar{\A} \right)^T}. \nonumber
\end{align}

Therefore, the total prior variance follows from \eqn{cond:var}, \eqn{totalStatVar}, and \eqn{var_mu} as

\begin{align}
& \sigb^{-}_{k}  := \E{\sigb^{-}_{k}(\param_{k-1})} + \textbf{Var}\left(\mub^{-}_{k}(\param_{k-1})\right) \eqnlabel{var_prior}  \\
&=\E{ \A(\param_{k-1})\sigb^{+}_{k-1}\A^T(\param_{k-1})} \nonumber \\ &+ \E{\B(\param_{k-1}) \Q \B^T(\param_{k-1}) }\nonumber \\
&+ \Exp{\left(\A(\param_{k-1})-\bar{\A}\right) \left(\mub^{+}_{{k-1}}  \mub^{+}_{k-1}{}^T \right) \left(\A(\param_{k-1})-\bar{\A}\right)^T}. \nonumber
% &+ \Exp{\Ap\mub^{+}_{{k-1}} ( \mub^{+}_{{k-1}})^T \Ap^T} - \bar{\A} \mub^{+}_{{k-1}}( \mub^{+}_{{k-1}})^T \bar{\A}^T
\end{align}

\subsubsection{\textbf{Update}}\label{sec:update}
Since we have assumed the matrix $\C$ in the measurement model \eqn{y_dt} to be independent of $\param_{k}$, we can simply follow the standard Kalman update equations. For the brevity of discussion, we omit the step by step derivation of the well known Kalman gain and update equations, which can be found in many textbooks, e.g. \cite{BrysonBook}. Once we have the propagated priors from equations \eqn{mean_prior} and  \eqn{var_prior}, the posteriors are given by
\begin{subequations}
\begin{align}
 \mub^{+}_{k} &= \mub^{-}_{k} + \K_k\left(\y_k - \C\mub^{-}_{k}\right), \eqnlabel{mean:update1} \\
\sigb^{+}_{k} &= \left(\vo{I}-\K_k \C \right)\sigb^{-}_{k}, \eqnlabel{var:update1}
\end{align}
\eqnlabel{disc_update1}
\end{subequations}
where $\y_k$ is the sensor measurement, and $\K_k :=\sigb^{-}_{k}\C^T$ $[\C\sigb^{-}_{k}\C^T+\R]^{-1}$ is the optimal Kalman gain. % to achieve an unbiased estimate with the minimum state error variance.

\subsection{Continuous-Discrete Robust Kalman Filter}\label{sec:CRF}
The continuous-discrete filter, also known as the hybrid Kalman filter, is more practical than other filters as it is suitable for most physical dynamical systems that are governed by continuous time ODEs, and sensor measurements are available only at discrete time instants. The system and sensor equations follow from \eqn{sys_ct0} and \eqn{y0} {for $t \in [t_{k-1}, t_k)$,}
\begin{subequations}
\begin{align}
\xdot(t) &= {\A(\param_{k-1}) \x(t) + \B(\param_{k-1})\w(t),} \elab{sys_ct} \\
\y(t_k) &= \C\x(t_k) + \n(t_k) .\elab{y_ct}
\end{align}
{Hereafter, for notational convenience, we drop the subscript $k-1$, and denote $\param_{k-1}$ by $\param$, since it does not vary in the interval $[t_{k-1}, t_k)$.}
\end{subequations}
\subsubsection{\textbf{Uncertainty Propagation}}\label{sec:UP}
Determining the moments of $\x(t,\param,\w)$, {the solution of \eqn{sys_ct}}, is nontrivial in this case, particularly due to $\param$. This can be shown by first defining mean and covariance as
\begin{align*}
\mub(t) &:= \Exp{\x(t,\param,\w)}, \text{ and }\\
\sigb(t) &:= \Exp{\left(\x(t,\param,\w)-\mub(t)\right)\left(\x(t,\param,\w)-\mub(t)\right)^T}.
\end{align*}
The propagation equation for $\mub(t)$ is given by
% $$\dot{\mub}(t) = \E{\A(\param)\x(\param,t)} + \E{\B(\param)\w(t)},$$
{$$\dot{\mub}(t) = \E{\A(\param)\x(t)} + \E{\B(\param)\w(t)},$$}
which presents a challenge in solving the differential equation due to uncertain matrices $\Ap$ and $\Bp$. Similar difficulty is faced in the propagation equation for $\sigb(t)$. We next present an approach based on the polynomial chaos theory to determine the first two moments of $\x(t,\param,\w)$.

As in the previous section, we adopt the formulation based on the conditional expectation with respect to $\param$. For a given $\param$, we can write the propagation equation for conditional mean and variance as
\begin{subequations}
\begin{align}
\dot{\mub}(t,\param) &= \Ap\mub(t,\param), \eqnlabel{cond:muCT}\\ %\elab{cond:mu}\\
\dot{\sigb}(t,\param) &= \Ap\sigb(t,\param)  + \sigb(t,\param)\Apt \nonumber \\ & \quad \quad + \Bp\Q\Bpt, \eqnlabel{cond:varCT}
\end{align}
\end{subequations}

The total mean and variance of {$\x(t,\param,\w)$} can be computed as
\begin{subequations}
\begin{align}
\mub(t) &:= \E{\mub(t,\param)},\\
\sigb(t) &:= \E{\sigb(t,\param)}  + \textbf{Var}(\mub(t,\param)). % \Exp{\left(\mub(t,\param)-\mub(t)\right)\left(\mub(t,\param)-\mub(t)\right)^T}.
\end{align}
\eqnlabel{totalStatCT}
\end{subequations}
Stochastic processes $\mub(t,\param)$ and $\sigb(t,\param)$ are expanded with polynomial chaos basis functions as follows.
\\
\underline{\textbf{Polynomial chaos expansions:}}
The expansion for $\mub(t,\param)$ follows from \eqn{compactxpc} as
\begin{align}
\hat{\mub}(t,\param) &= \sum_{i=0}^{N} \mub_i(t)\phi_i(\param)
=\begin{bmatrix} \mub_0(t) & \cdots & \mub_N(t)\end{bmatrix}\mo{\Phi}(\param) \nonumber\\
&=  \vo{{\tilde{\mu}_{\text{pc}}}}\Phin{n} = \Phint{n}\mub_{\text{pc}} , \eqnlabel{mu_pc}
%\vo{\widetilde{\mu_{\text{pc}}}}\Phin{n} \\
% &= (\mo{\Phi}^{T}(\param)\otimes \I{n}) \vo{\mub}_{\text{pc}}
\end{align}
where, $\vo{{\tilde{\mu}_{\text{pc}}}} := \begin{bmatrix} \mub_0(t) & \cdots & \mub_N(t)\end{bmatrix} \in \real^{n\times(N+1)}$, and $\mub_{\text{pc}} := \vec{\vo{{\tilde{\mu}_{\text{pc}}}}} \in \real^{n(N+1)}$.

Since $\sigb(t,\param)\geq \vo{0}$, the stochastic process $\sigb(t,\param)$ is expanded using quadratic basis functions constructed from $\phi_i$. We adopt the expansion presented in
\cite{bhattacharyarobust}, i.e.
\begin{align}
\hat{\sigb}(t,\param) &= \Phint{n} \begin{bmatrix} \sigb_{00} & \cdots & \sigb_{0N} \\ \vdots &  & \vdots \\ \sigb_{N0}  & \cdots & \sigb_{NN}
\end{bmatrix} \Phin{n}. \nonumber
\end{align}
Since $\hat{\sigb}(t,\param)$ is symmetric and $\hat{\sigb}(t,\param)\geq0$ , it follows that $\sigb_{ij} = \sigb^T_{ij} = \sigb_{ji} \geq 0$. Therefore, $\hat{\sigb}(t,\param)$ can be expanded as
\begin{align}
\hat{\sigb}(t,\param) &=  \sum_{ij} \sigb_{ij}(t)\phi_i(\param)\phi_j(\param). \nonumber
\end{align}
Moreover, we note that the quadratic basis functions, $\{\phi_i(\param)\phi_j(\param)\}$, are not linearly independent. Therefore, the PC expansion for $\hat{\sigb}(t,\param)$ can be effectively written as
\begin{align}
\hat{\sigb}(t,\param) &=  \sum_{i=0}^M \sigb_{i}(t)\theta_i(\param) = \big[\sigb_{0}(t),\cdots, \sigb_{M}(t) \big] \Thp  \nonumber \\ &=  (\Thpt \otimes \vo{I}_n)\sigb_{\text{pc}} =: \Thnpt{n}\sigb_{\text{pc}}, \eqnlabel{var_pc}
\end{align}
where, $M:=2(N-1)$,  $0\leq\sigb_{i}(t) \in \real^{n\times n}$ and
\begin{align}
\sigb_{\text{pc}} &:=
\begin{bmatrix} \sigb_0(t) & \cdots & \sigb_M(t)\end{bmatrix}^T \in \real^{n(M+1)\times n}.\nonumber
\end{align}
The basis functions $\theta_i(\param)$ are linearly independent polynomials chosen from quadratic terms resulting from the expansion of $\big(\phi_0(\param)+\phi_1(\param)+\cdots+\phi_N(\param)\big)^2$, i.e.  $\theta_i(\param)$ are linearly independent basis functions selected from the following set
\begin{align*}
\begin{Bmatrix} \phi_0(\param)\phi_0(\param) \\ 2\phi_0(\param)\phi_1(\param) \\ \vdots\\ 2\phi_{N-1}(\param)\phi_{N}(\param) \\ \phi_{N}(\param)\phi_{N}(\param)\end{Bmatrix},
\end{align*}
and, $\Thp := \left[ \theta_0(\param), \theta_1(\param), \cdots, \theta_M(\param)\right]^T \in \real^{M+1}$.
With this mean and variance approximation, the error equations in \eqn{cond:muCT} and \eqn{cond:varCT} are
\begin{subequations}
\begin{align}
&\vo{e}_{\mub}(t,\param) := \Phint{n}\dot{\mub}_{\text{pc}} - \Ap\Phint{n}\mub_{\text{pc}}, \text{ and }\\
&\vo{e}_{\sigb}(t,\param) := \Thnpt{n}\dot{\sigb}_{\text{pc}} - \Ap\Thnpt{n}\sigb_{\text{pc}} \nonumber \\
 & \quad \quad \quad -\Thnpt{n}\sigb_{\text{pc}} \Apt - \Bp\Q\Bpt.
\end{align}
\eqnlabel{eqnerror}
\end{subequations}
The differential equations for $\dot{\mub}_i(t)$ and $\dot{\sigb}_i(t)$ are obtained by setting
\begin{align*}
\E{\vo{e}_{\mub}(t,\param)\phi_i(\param)} = 0, \text{ and }
\E{\vo{e}_{\sigb}(t,\param)\theta_j(\param)} = 0,
\end{align*}
for $i=0,\cdots,N$; and $j=0,\cdots,M$,

resulting in
\begin{align}
 \dot{\mub}_{\text{pc}}  = \A_{\mub}  \mub_{\text{pc}}\, , \, \text{ and } \,
 \dot{\sigb}_{\text{pc}} = \F_{\sigb}  + \B_{\sigb}\, , \text{ where, }
\eqnlabel{propagation}
\end{align}
\begin{align*}
&\A_{\mub} :=\E{\Phin{n}\Phint{n}}^{-1}\E{\Phin{n}\Ap\Phint{n}},  \\
&\F_{\sigb} := \E{\Thnp{n}\Thnpt{n}}^{-1}  \times \\ & \E{\Thnp{n}\Ap\Thnpt{n}\sigb_{\text{pc}}
+ \Thnp{n}\Thnpt{n}\sigb_{\text{pc}}\Apt},  \\
&\B_{\sigb} := \E{\Thnp{n}\Thnpt{n}} ^{-1} \E{\Thnp{n}\Bp\Q\Bpt}.
\end{align*}

\noindent \underline{\textbf{Computation of the prior:} }
Given the posteriors $\mub^+(t_{k-1})$ and $\sigb^+(t_{k-1})$, at time instant $t_{k-1}$, the evolution of the state uncertainty is determined by integrating \eqn{propagation} over $[t_{k-1},t_k]$ to arrive at $\mub^-(t_k,\param)$ and $\sigb^-(t_k,\param)$, the conditional prior mean and the conditional prior variance of the state. The total mean and covariance priors, i.e. $\mub^-(t_k)$ and $\sigb^-(t_k)$, are then determined from \eqn{totalStatCT}.

Integration of \eqn{propagation} requires initial conditions $\mub_\text{pc}^+(t_{k-1})$ and $\sigb_\text{pc}^+(t_{k-1})$, which are determined by projecting $\mub^+(t_{k-1})$ and $\sigb^+(t_{k-1})$ on the basis functions $\{\phi_i(\param)\}_{i=0}^{N}$, and $\{\theta_i(\param)\}_{i=0}^{M}$ respectively. Noting that $\mub^+(t_{k-1})$ and $\sigb^+(t_{k-1})$ are $\param$ independent, initial conditions $\mub_\text{pc}(t_{k-1})$ and $\sigb_\text{pc}(t_{k-1})$ are given by
\begin{align}
\mub_\text{pc}^+(t_{k-1}) := \begin{bmatrix}
\mub^+(t_{k-1}) \\
\vo{0}_{nN}
\end{bmatrix},
\sigb_\text{pc}^+(t_{k-1}) := \begin{bmatrix}
\sigb^+(t_{k-1}) \\
\vo{0}_{nM\times n}
\end{bmatrix}. \nonumber
\end{align}
With these initial conditions, linear ODEs \eqn{propagation} can be integrated to calculate $\mub_\text{pc}^-(t_{k})$ and $\sigb_\text{pc}^-(t_{k})$ at time $t_k$.

Therefore, conditional mean and covariance priors at $t_k$ follow from \eqn{mu_pc} and \eqn{var_pc} as  \begin{align*}\mub^-(t_k,\param) &= \Phint{n}\mub_\text{pc}^-(t_{k}), \\ \sigb^-(t_k,\param) &= \Thnpt{n}\sigb_\text{pc}^-(t_{k}).\end{align*}
The total mean and covariance priors $\mub^-(t_k)$ and $\sigb^-(t_k)$ are calculated using \eqn{totalStatCT} as follows.
\begin{align}
\mub^-(t_k) &= \E{\Phint{n}} \mub_\text{pc}^-(t_{k}), \nonumber \\
\sigb^-(t_k) &= \E{\Thnpt{n}}\sigb_\text{pc}^-(t_{k}) \nonumber  \\
&+ \vo{\tilde{\mu}^-_{\text{pc}}}
(t_{k}) \Big(\textbf{Var}\left(\mo{\Phi}(\param)\right) \Big) \big( \vo{\tilde{\mu}^-_{\text{pc}}}(t_{k})\big)^T, \nonumber
\end{align}
where, $\textbf{Var}\left(\mo{\Phi}(\param)\right):= \mathbb{E}\Big[\big(\mo{\Phi}(\param) - \overline{\vo{\Phi}} \big) \big(\mo{\Phi}(\param)- \overline{\vo{\Phi}} \big)^T  \Big]$, and $\overline{\vo{\Phi}}:=\E{\mo{\Phi}(\param)}$.

\subsubsection{\textbf{Update}}
Since the measurements are obtained at discrete time instants, we can use the Kalman update equations from \S \ref{sec:update}.
The updated posteriors are given by
\begin{subequations}
\begin{align}
 &\mub^{+}(t_k) = \mub^{-}(t_k) + \K_k\left(\y(t_k) - \C\mub^{-}(t_k)\right), \nonumber \\
&\sigb^{+}(t_k) = \big(\vo{I}-\K(t_k) \C \big)\sigb^{-}(t_k), \nonumber
\end{align}
\end{subequations}
where, $\y(t_k)$ is the sensor measurement, and $$\K(t_k):=\sigb^{-}(t_k)\C^T[\C\sigb^{-}(t_k)\C^T+\R]^{-1}.$$

\section{Numerical Results}\label{sec:result}
Performance of the proposed robust Kalman filter is tested with two cases of simulation: 1) Case I: Initial mean, $\mu_0 = [0 \ 0]^T $, for checking steady state error, 2) Case II: Initial mean, $\mu_0 \neq [0 \ 0]^T $, for checking convergence rate with initial uncertainty.% which is the indicator for how much filter rapidly act on input signal.
  We compare the performance of the filter in terms of the estimation accuracy characterized by the mean and standard deviation (SD) of absolute error, and the rate of convergence.
\subsection{Discrete Robust Kalman filter}
The proposed discrete robust Kalman filter discussed in \S \ref{sec:DRF} is applied to the example (\ref{eqn:example1}) that was previously considered as a test problem in \cite{xie1994robustDT,zhu2002design}.
\begin{subequations}
\begin{align}
\x_{k} &= \begin{bmatrix}0 & -0.5 \\ 1 &1 + \delta\end{bmatrix} \x_{k-1} +\begin{bmatrix} -6 \\ 1 \end{bmatrix}\w_{k-1}, \\
\y_k &= \begin{bmatrix} -100 & 10\end{bmatrix}\x_{k} + \n_k .
\end{align}\label{eqn:example1}
\end{subequations}
where $\delta$ is a uniformly distributed random parameter in $[-0.3 \ 0.3]$, and the variance of process and measurement noise is assumed to be unity, i.e. $Q=1$, $R=1$.

We choose uniformly spaced $10$ points in $[-0.3 \ 0.3]$ as samples for $\delta$. Then, mean and standard deviation of absolute error obtained for different realizations of the plant corresponding to different values of $\delta$, are considered as metrics for the estimation accuracy. We compare the performance of the proposed filter with standard Kalman filter with nominal plant realization corresponding to $\delta = 0$.
As claimed by the authors of \cite{zhu2002design}, and verified by us, the filter presented in \cite{zhu2002design} performs better than the one discussed in \cite{xie1994robustDT}. Therefore, herein, we compare the performance of the proposed filter only with \cite{zhu2002design}.
%We also consider the robust filters presented in \cite{zhu2002design, xie1994robustDT} for comparison of results.
%\comment{We also consider the robust filter presented in \cite{zhu2002design} which performs  better than the one presented in \cite{xie1994robustDT}, for comparison of results. -- there is only one robust filter being compared to.}

\begin{figure}[htb!]
\centering
\includegraphics[width=0.48\textwidth]{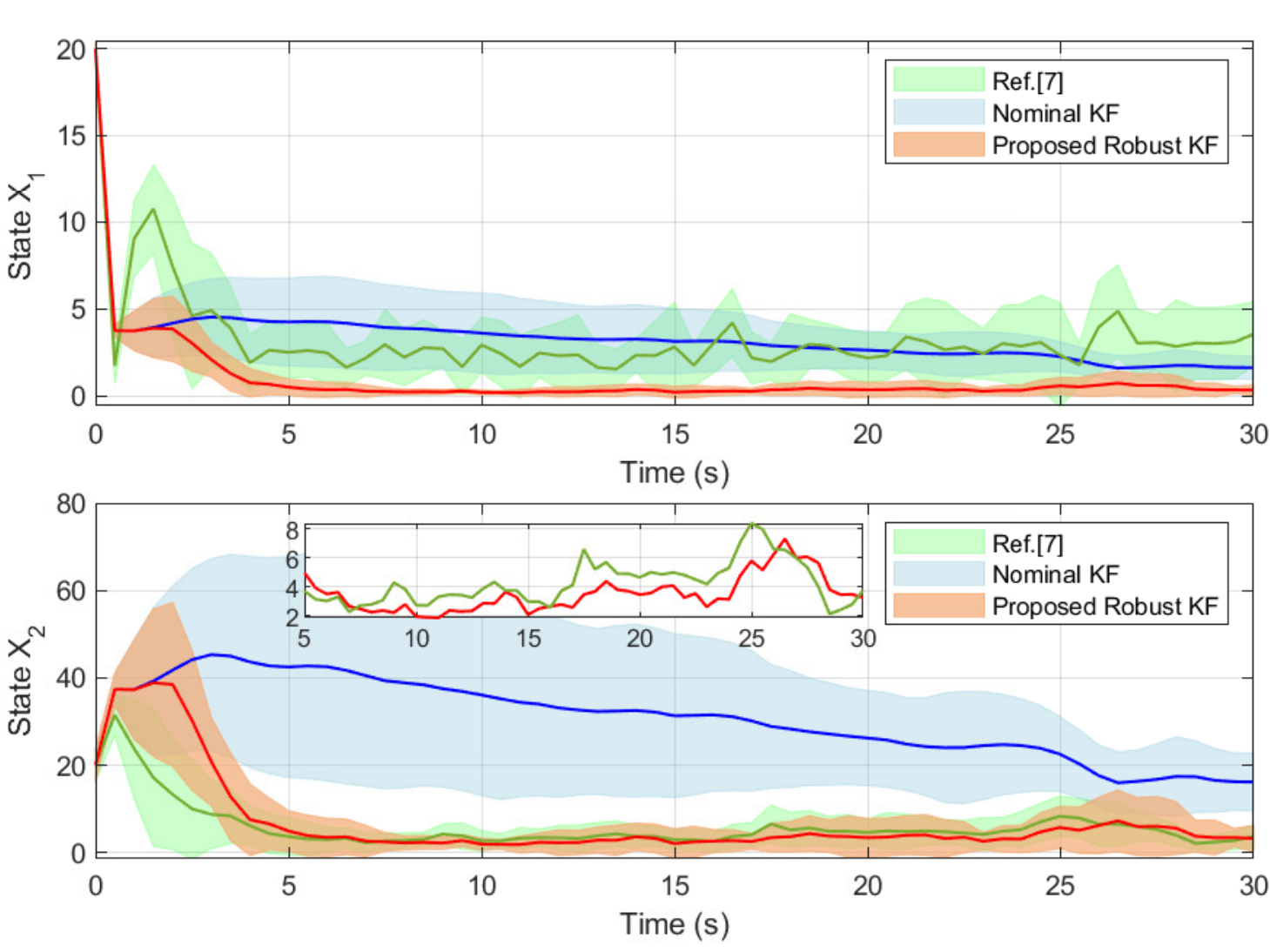}
\caption{DT robust filter: Mean and standard deviation of the absolute error with initial condition $x_0 = [ 20\ 20]^T$, for Case II. }
\label{fig:DT_2}
\end{figure}

The simulation results for the proposed discrete robust Kalman filter, the nominal Kalman filter, and the filter from \cite{zhu2002design}, are shown in  Fig. \ref{fig:DT_2} and TABLE \ref{Table:RMS_DT}.
In both simulation Cases I and II, the proposed robust Kalman filter has the least mean error than the other filters, as shown in TABLE \ref{Table:RMS_DT}.
Moreover, for Case II as shown in Fig. \ref{fig:DT_2}, the proposed filter converges faster than the nominal KF, and its convergence rate is comparable to the filter from \cite{zhu2002design}. We also note that the computational time required for the proposed filter is comparable to that of nominal KF and the filter from \cite{zhu2002design}.

\begin{table}[t]
\caption{Comparison of error in Discrete time filters.}
\vspace{-0.4cm}
\label{Table:RMS_DT}
\begin{center}
\renewcommand{\arraystretch}{1.5}
\begin{tabular}{|c|c||c|c|c|}
\hline
\multicolumn{2}{|c||}{\multirow{2}{*}{Filter Algorithm}} & Case I &  Case II \\
\multicolumn{2}{|c||}{} & Mean \ / \ SD & Mean \ / \ SD \\
\hline \hline
% \multirow{2}{5em}{Ref.\cite{xie1994robustDT}} & $x_1$ & 3.2256 / 2.1266 &   4.0866 / 2.4553   \\
                        %   & $x_2$ & 5.7093 / 4.1997 &   8.1330 / 4.9630   \\
% \hline
\multirow{2}{5em}{Ref.\cite{zhu2002design}} & $x_1$ & 2.7325 / 1.7518 &   3.4675 / 2.0216   \\
                            & $x_2$ & 4.3049 / 2.9640 &   6.0837 / \textbf{3.5487}   \\
\hline
\multirow{2}{5em}{Nominal KF} & $x_1$ & 0.4438 / 0.4136  & 2.4085 / 1.0595 \\
                              & $x_2$ & 4.4418 / 4.1355 & 24.0821 / 10.5982 \\
\hline
\multirow{2}{5em}{Proposed Robust KF} & $x_1$ & \textbf{0.3182} / \textbf{0.2914} &   \textbf{0.5666} / \textbf{0.4314}  \\
                                      & $x_2$ & \textbf{3.1846} / \textbf{2.9114} &   \textbf{5.6669} / 4.3099   \\
\hline
\end{tabular}
\end{center}
\vspace{-0.6cm}
\end{table}
\subsection{Continuous-Discrete Robust Kalman Filter}
The proposed hybrid robust Kalman filter in \S \ref{sec:CRF} is applied to the example (\ref{eqn:example2}) and its performance is compared with the nominal Kalman filter.
% Our example for the hybrid robust Kalman filter is
\begin{subequations}
\begin{align}
 \xdot(t) &= \begin{bmatrix} 0 & -1 + \delta \\ 1 & -0.5 \end{bmatrix} \x(t) +\begin{bmatrix} -2 \\ 1 \end{bmatrix}\w(t), \\
\y(t_k) &= \begin{bmatrix} -100 & -100\end{bmatrix}\x(t_k) + \n(t_k) ,
\end{align}\label{eqn:example2}
\end{subequations}
where $\delta$ is uniformly distributed in the interval $[-0.95 \ 0.95]$, and the variances of process and measurement noise are $Q=1$ and $R=1$. We use the similar performance metrics discussed in the previous subsection.

\begin{figure}[htb]
\centering
\includegraphics[width=0.48\textwidth]{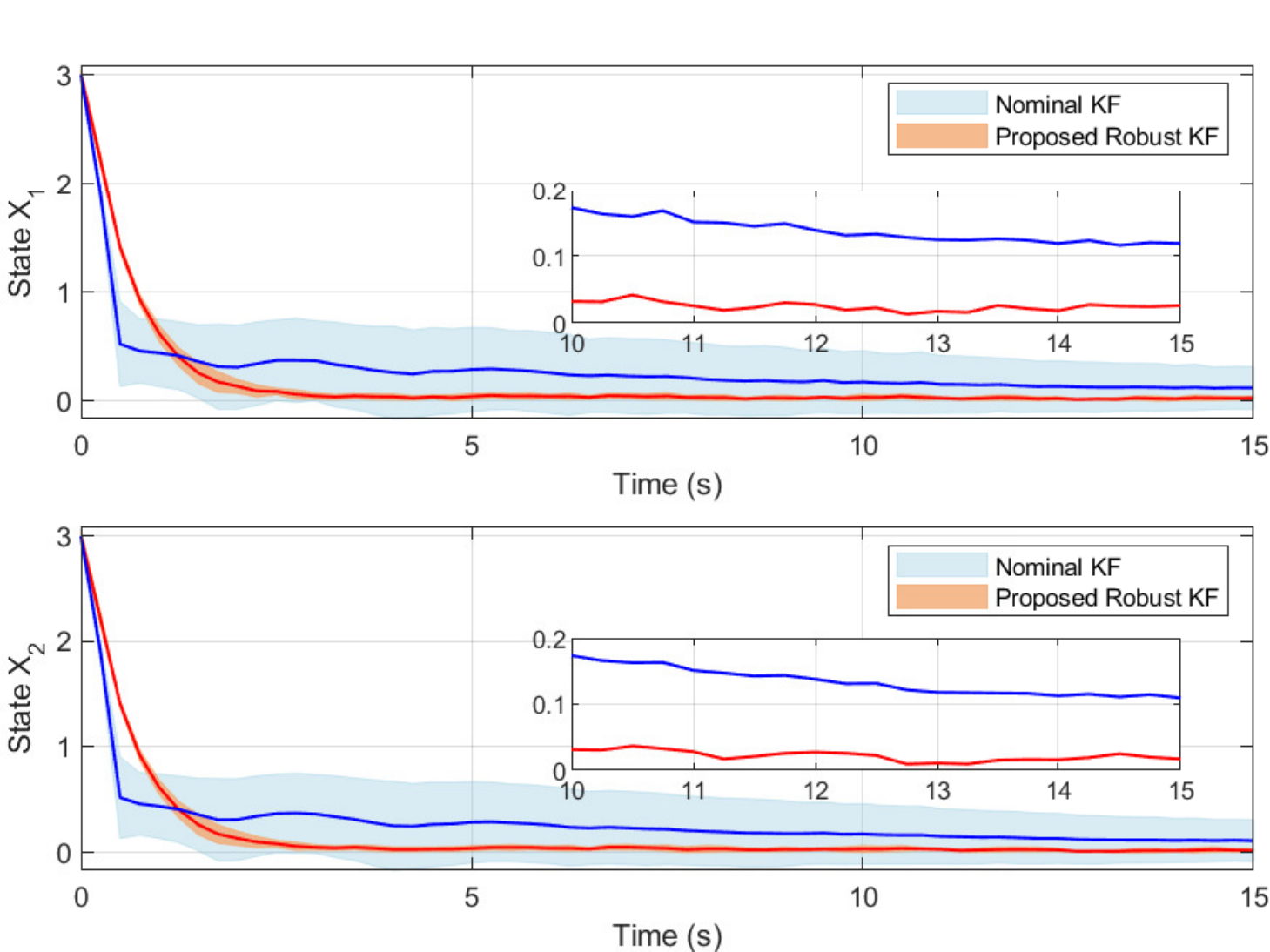}
\caption{Hybrid robust filter: Mean and standard deviation of the absolute error with initial condition $x_0 = [ 3\ 3]^T$, for Case II.}
\label{fig:CT2}
\end{figure}

The proposed robust Kalman filter is 2 times more accurate than the nominal KF in steady state as shown in TABLE \ref{Table:RMS_Hybrid}. Moreover, it shows faster convergence than the nominal KF as shown in Fig.\ref{fig:CT2}. Again, we note that the computational time required for the proposed filter is comparable to the nominal KF.

\begin{table}[htb]
\vspace{-0.5cm}
\caption{Comparison of error in hybrid filters.}\label{Table:RMS_Hybrid}
\vspace{-0.4cm}
\begin{center}
\renewcommand{\arraystretch}{1.5}
\begin{tabular}{|c|c||c|c|c|}
\hline
\multicolumn{2}{|c||}{\multirow{2}{*}{Filter Algorithm}} & Case I &  Case II \\
\multicolumn{2}{|c||}{} & Mean \ / \ SD & Mean \ / \ SD \\
\hline \hline
\multirow{2}{5em}{Nominal KF} & $x_1$ & 0.0223 / 0.0204  & 0.2052 / 0.1694\\
& $x_2$ & 0.0195 / 0.0211 & 0.2038 / 0.1695 \\
\hline
\multirow{2}{5em}{Proposed Robust KF } & $x_1$ & \textbf{0.0155} / \textbf{0.0092} &   \textbf{0.1833 }/ \textbf{0.0783}   \\
& $x_2$ & \textbf{0.0137} / \textbf{0.0077} &   \textbf{0.1822} / \textbf{0.0782}   \\
\hline
\end{tabular}
\end{center}
\vspace{-0.6cm}
\end{table}

\section{Conclusion}\label{sec:Conclusion}
In this paper, we proposed robust Kalman filter with probabilistic uncertainty in system parameters. Mean and variance of the uncertain system are propagated using conditional probability and the polynomial chaos (PC) expansion framework.
{The empirical results in this preliminary work show that the proposed approach which exploits the information about probability distribution of the uncertain parameters, demonstrates better performance than the existing frameworks which are designed for the worst case scenarios that occur with the vanishing probability. This serves as a motivation to pursue a theoretical treatment of the performance guarantees for the proposed approach, which is a topic of our ongoing research.}

\bibliographystyle{unsrt}
\bibliography{robustEstimation}

\end{document}